\documentclass[11pt]{article}
\usepackage{amssymb, epsfig}
\usepackage{multicol}
\newcommand{\newc}{\newcommand}
\newc{\beq}{\begin{equation}}
\newc{\eeq}{\end{equation}}
\newc{\beqa}{\begin{eqnarray}}
\newc{\eeqa}{\end{eqnarray}}
\newc{\IM}{\mbox{\sl{Im}}}
\newc{\RE}{\mbox{\sl{Re}}}
\newc{\nonr}{\nonumber}
\newc{\hs}{\hskip 3mm}
\newc{\ra}{\rightarrow}
\newc{\TR}{\mbox{\sl{Tr}}}
\newc{\tri}{\triangle}
\newc{\trip}{$\mathbf{3}$ }
\newc{\sext}{$\bar{\mathbf{6}}$ }
\newc{\tripp}{$\mathbf{3^{\prime}}$ }
\newc{\frpir}{\frac{1}{\sqrt { \pi R}}}
\newc{\bitem}{\begin{itemize}}
\newc{\eitem}{\end{itemize}}
\newc{\bcn}{\begin{center}}
\newc{\ecn}{\end{center}}
\newc{\etal}{{\it {et al}}}
\newc{\bray}{\begin{array}}
\newc{\eray}{\end{array}}
\newc{\bdpm}{\begin{displaymath}}
\newc{\edpm}{\end{displaymath}}
\newc{\benu}{\begin{enumerate}}
\newc{\eenu}{\end{enumerate}}

\newc{\PRD}[3]{{Phys.~Rev.} \textbf{D#1},({#2}) #3}
\newc{\PLB}[3]{{Phys.~Lett.} \textbf{B#1}, ({#2}) #3}
\newc{\PRL}[3]{{Phys.~Rev.~Lett.} \textbf{#1}, ({#2}) #3}
\newc{\NPB}[3]{{Nucl.~Phys.} \textbf{B#1}, ({#2}) #3}
\newc{\NPA}[3]{{Nucl.~Phys.} \textbf{A#1}, ({#2}) #3}
\newc{\NPPS}[3]{{Nucl.~Phys.~Proc.~Suppl} \textbf{#1},({#2}) #3}
\newc{\PTP}[3]{{Prog.~Theo.~Phys.} \textbf{#1}, ({#2}) #3}
\newc{\AJS}[3]{{Astrophysical~J.~Suppl} \textbf{#1}({#2}) #3}
\newc{\JHEP}[3]{{JHEP}\textbf{#1}({#2}) #3}
\begin{document}
\begin{center}
{\bf Neutrino Mass Models in Extra Dimensions}\\
\vspace{1cm}
John N.Ng\\
Triumf Theory Group, 4004 Wesbrook Mall\\
Vancouver, B.C.\\
Canada V6T 2A3\\
\end{center}
\begin{abstract}
Neutrinos play a crucial role in many areas of physics from particle physics at very
short distances  
to astrophysics and cosmology. It is a long held believe that they are good probes of physics
at the GUT scale. Recent developments have made it clear
that they can also be of fundamental importance for the physics of extra dimensions
if these exist. Here we  pedagogically review the construction of neutrino mass models 
in extra dimensions within the brane scenarios.
These models are usually nontrivial generalization of  their four dimensional counter parts. We
describe the theoretical tools that have been forged and the new perspectives gained in
this rapidly developing area.
In particular we discuss the issues
involve with building models without the use of right-handed singlets. It is very 
difficult to directly test the origin of neutrino masses in the different models be it in
four or more five dimensions. We point out that different  
 models give very different indirect signatures at the TeV region and precision measurements. 
\end{abstract}
\section{Introduction}
We have now convincing evidence that the three active neutrinos of the standard model have different
masses and they mix with each other. The results of the SuperK\cite{SuperK}, SNO\cite{SNO}
 and Kamland \cite{Kamland}
experiments have now 
narrow the $\nu$ mass pattern to one of three possibilities
\bitem
\item Inverted mass hierarchy (IMH)
\item Normal mass hierarchy (NMH)
\item Almost degenerate masses
\eitem
The first two cases are depicted in Fig. 1 and we concentrate on them.
\begin{figure}[h]
\bcn
\epsfig{file=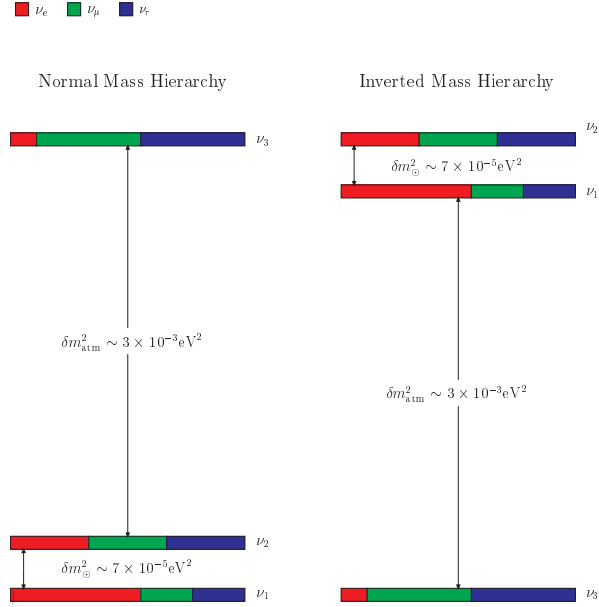,height=8cm}
\caption{Neutrino Mass Patterns}
\label{fig:numass}
\ecn
\end{figure}
The notations in the figure are standard.
However, the overall mass scale is not known. We only have upper bounds of $< 1$eV from
WMAP \cite{WMAP} and $<2.2$eV  from tritium end point experiments \cite{TM}. 
If neutrinos have Majorana masses
then a better limit of $< .3$ eV comes from neutrinoless double beta decay $(\beta \beta)_{0\nu}$
experiments \cite{0nu}. 
This latter result is model dependent and is subject to uncertainties in nuclear matrix
element calculations as well as the implicit assumption that neutrino mass is the dominant 
effect leading to the decay. There are many excellent reviews on these are
other questions we list the most recent ones in \cite{0nurev}. 

As with other fermions when neutrinos are massive their mass eigenstates and weak eigenstates 
do not coincide. In general they are related via a mixing matrix  $U_{PMNS}$ \cite{PMNS}.
Analysis of the data yield 
\beqa
 U_{PMNS}&=& \left( \begin{array}{ccc} U_{e1}&U_{e2}&U_{e3}\\U_{\mu 1}&U_{\mu 2}&U_{\mu 3}\\
U_{\tau 1}&U_{\tau 2}&U_{\tau 3} \end{array} \right) \nonumber\\
&=&\left( \begin{array}{ccc}0.79-0.86&0.50-0.61&.0-0.16\\ 0.24-0.52&0.44-0.69&0.63-0.79\\
0.26-0.52&0.47-0.71&0.60-0.77 \end{array} \right).
\label{eq:MNS}
\eeqa
In a gauge theory $U_{PMNS}$ is a product of two mixing matrices :
\bdpm
U_{PMNS}=U_l U_{\nu}
\edpm
where $U_l$ diagonalizes the charged lepton mass matrix and $U_{\nu}$ does the same
for neutrinos. Often one chooses the mass eigenbasis for the charged leptons and $U_{PMNS}$
is just the neutrino mixing matrix. 
Furthermore, Eq.(\ref{eq:MNS}) is a low energy solution. It is typical that gauge models
are formulated at some high energy scale and  reach the above via renormalization group running . 
Whether this
has a large effect is highly model dependent. In most of our discussions we assume that 
this is not large. 

Now we return  to the mixing matrix.  It is 
clear that, unlike the down quark mixing, neutrino mixing are  bilarge and the recent
SNO data also rule out bi-maximal. One can reconstruct the neutrino mass matrix in
the weak basis ${ \cal{M}}_{\nu}$ via ${\cal{M}}_{\nu}={U^T}_{PMNS}M_DU_{PMNS}$ where $M_D$ is the
diagonal mass matrix with eigenvalues $(m_1, \ m_2, \ m_3)$. It is ${\cal M}_{\nu}$ that is
of the greatest interest to theorists. In general $m_{1,2,3}$ are
complex numbers for Majorana neutrinos. Customarily one phase is put into $U_{PMNS}$ and
 two are left in $M_D$. With the values of mass differences depicted in Fig. 1 and
Eq. (\ref{eq:MNS}) the leading mass patterns are given by 
\benu
\item IMH
\benu
\item
\beq
{\cal{M}}_{\nu} \sim m_0\left( \bray{ccc}1&\times&\times\\ \times&1/2&1/2\\ \times&1/2&1/2
\eray \right)
\eeq
\item
\beq
{\cal{M}}_{\nu} \sim m_0\left( \bray{ccc}\times&1&1\\ 1 &\times&\times \\ 1&\times &\times
\eray \right)
\eeq
\eenu
\item NMH
\beq
{\cal {M}}_{\nu} \sim m_0 \left( \bray{ccc}\times&\times&\times \\ \times&1/2&-1/2 \\
\times& -1/2&1/2 \eray \right)
\eeq
\eenu
where $m_0$ is the unknown overall mass scale and 
$\times$ denotes some small number. The challenge to theorists is to
construct viable models that
give rise to one of the above structures. At the same time one gets a more or less
natural understanding of why the overall neutrino mass scale is so much smaller 
than the charged fermions;
i.e. what physics sets the scale of neutrino masses.

\section{Are Neutrinos Dirac or Majorana Particles?}

The strong evidence of neutrino oscillations and mixing has certainly shattered the
concept of separately conserved electron, muon or tau lepton number; however,
the question of whether total lepton number is conserved  is still open. The usual way 
of stating this is in terms of whether neutrinos are Dirac or Majorana particle.
If neutrinos were  Majorana then they are own anti-particle and lepton number 
violating reactions can be expected to proceed. Otherwise they are Dirac particles.
 Experimentally this is a very
difficult question to answer. The favorite process is  $(\beta \beta)_{0\nu}$ decays.
If one such decay  were observed then
it will be clear evidence that lepton number is not conserved and it will natural
to conclude that neutrinos are Majorana or at least a Majorana mass term is
not forbidden. On the other hand to prove that neutrinos are Dirac particles will be 
very difficult.

The nature of neutrinos has deep theoretical significance is modern particle
physics. The active neutrino in the SM is a left-handed Weyl particle; i.e. is a helicity
eigenstate. Due to gauge invariance it has no bare mass term. To give it a mass
one usually introduces a  right-handed fermion $N_R$ which is a SM singlet. Whether 
the  $N_R$ exists and what is its nature are two of the most 
fundamental question in neutrino physics. Certainly if $N_R$ exists how many are there?
What is its mass?  Is it  Dirac or Majorana particle?   
In the SM this is put in by hand and so the same is true $SU(5)$ GUT models. It is customary
to use at least one to generate masses for some of the active neutrinos 
and three if we believe in some family symmetry. In $SO(10)$ GUT
models $N_R$ naturally exists since the fundamental representation has 16 fermions
which is just the right number to accommodate each SM family plus a $N_R$. Here it
is natural to have three of them.

There are many theoretical reasons to pursue GUT models and understand neutrino
masses in this context (see \cite{King} for an up to date review); moreover, we
wish to study this in a wider setting. As is well known  $N_R$ being a singlet can have  a 
Majorana mass term , $(N_R)^cN_R$ as well as a Dirac mass term coupling given by
 $y\bar{\nu_L} N_R H^0$ where $ H^0$ is  the SM Higgs boson and $y$ is the Yukawa coupling.
After symmetry breaking the neutrino masses for 
one family is given by
\beq
\left( \bray{cc} \bar{\nu_L^c} &\bar{N_R^c} \eray \right)\left( \bray{cc} m_{\nu}&m_D\\
{m_D}^T &M_R \eray \right)
\left( \bray{c}\nu_L\\N_R \eray \right)
\label{eq:nummatrix}
\eeq
For three families each entry in the above is a $3\times 3$ matrix.

In the SM $m_{\nu}=0$ and $m_D\sim$ GeV. If $N_R$ is a Dirac fermion then $M_R=0$ and
we have Dirac masses for the active neutrinos. This is the simplest
extension of the SM. In this case it is very difficult to
understand the smallness of the neutrino masses and  extreme fine tuning
of the Yukawa coupling $\sim 10^{-11}$ to get it below the experimental bound.
In GUT models it is natural to take 
$M_R \sim 10^{14}$ GeV and we have the seesaw mechanism for generating neutrino
masses.

\section{$N_R$ as a bulk fermion}

Recently new perspectives in neutrino physics arise from the so call brane world scenario.
In the simplest form this makes use of the possibility that there are more then 4 dimensions.
The number of extra dimensions $\delta$ is between 1 and 7 as hinted by string
theory. They are taken to be spatial and can be relatively large \cite{LED}
which enables one to solve the hierarchy problem.  Fields with SM
quantum numbers are confined on a $1+3$ hyper surface whereas SM singlets such
as the graviton can propagate in the full $4+\delta$ dimensional bulk. The extra dimensions
are compactified in tori of radii $R_i$ where $i=1,\dots 6$. Introducing $N_R$ as a bulk 
field was done in \cite{bulkn}. As is well known compactification leads to 4D Kaluza-Klein(KK)
excitations of $N_R$ with masses $n/R, n=0,1,2,\dots$ where we have taken 
all radii to be equal. The zero mode can now couple to the active neutrino on the
brane and the SM Higgs in the usual way. If $N_R$ is taken to be Dirac then  a small
Dirac neutrino mass can be generated given by
\beq
m_D=\frac{yv}{ (2 \pi RM_*)^{\delta/2}}    \;\;\;\;\;\; v=245 \mathrm {GeV}
\eeq
without excessive fine tuning of $y$. This is  due to the volume dilution 
factor of $RM_*$ where $M_*\sim 10 \mathrm {TeV} $ 
is the fundamental scale for the theory we can obtain $m_D \sim 10^{-4}$eV.
Even with this simple model and only one family there is an important consequence for
neutrino oscillation. The KK excitations form an infinite tower of 
sterile neutrinos. One would expect an effective active-sterile oscillation \cite{bulkosc}in the
Superk data. However, there is no evidence  in the data. This may indicate that 
the mixing between the active neutrinos and the KK modes are small 
and/or $R$ cannot be taken too large. Other 
 effects in nuclear and astrophysics of this
simple model are discussed in \cite{McLN}.

We have seen that  small Dirac neutrino masses are quite natural in extra dimensional model;
however, to obtain the hierarchy indicated by the data is more difficult. One way is to
make use of family symmetry as in 4D models. Another solution is purely  extra dimensional 
which can also incorporate the masses of quarks and charged leptons. This is the split fermion
scenario \cite{SF} where the left-handed and right-handed fields are localized at different positions
in the extra dimensional space. For example, the $(\nu_e e)_L$ doublet is at $z=0$ and $e_R$ 
is at $z=z_1$ where $z$ denotes the extra dimension. This is repeated for other 
fermions. The gauge and Higgs fields
propagate in the bulk. The chiral fermions are also bulk fields 
and their zero modes, denoted by $\Psi$, are given  Gaussian profiles in $z$. For a fermion 
located at $z_i$ its wave function is given by
\beq 
\Psi_i(x,z)\sim {1 \over
\pi^{\frac14}\sigma^{\frac12} }\Psi_i(x)
e^{-\frac{(z-z_i)^2}{2\sigma^2}}. 
\eeq
where $\sigma$ is a free parameter that is taken to be universal for all fermions for
simplicity. The  product of two
fermion fields can be approximately replaced by 
\beq
\overline{\Psi}_i(x,z)\Psi_j(x,z) \sim
\exp\left(-\frac{(\triangle_{ij})^2}{4\sigma^2}\right)
\delta(z-\bar{z}_{ij})\bar{\Psi}_i(x)\Psi_j(x) \eeq where
$\bar{z}_{ij}=(y_i+y_j)/2$ is their average positions and
$\triangle_{ij}=z_i-z_j$.
After integrating out $z$ it is the overlap of the two different chiralities
of a fermion that determines its mass. Explicitly it is given by $\triangle{ij}$. The origin
of the  large mass for fermion is due to the proximity of its left and right-handed components
in the extra dimension. Similarly, if the chiral components of a fermion is far apart
in $z$ it will be light.
Hence, one does 
not need to fine tune Yukawa couplings but instead the positions $z_i$ are used to fit the data.
It is found that  realistic quark and lepton mass matrices and an acceptable 
 CKM matrix can be
be found \cite{SFphen}. In this scenario the fermion masses becomes a geographical 
problem in the extra dimension. Stating it differently we need a 
mechanism that  puts the fermions in their correct positions instead of putting it in by hand. 
As a result, although it is
non-trivial to be able to reproduce the observed masses and mixing, these models
suffer from a lack of predictive power.

\section{ What if there is no $N_R$?} 
As can be seen from Esq.(\ref{eq:nummatrix}) that neutrinos will be massless 
in the SM and can have
only a Majoring mass via radiative effects. This is first done in \cite{Zee} 
within the SM gauger group by adding a $SI(2)$ singlet but $U(1)$ charged scalar field.
The model also requires an additional doublet Hogs field $S^+$. Due to Fermi statistics
$S$ can only couple to leptons of two different families. The Feynman diagram generating
neutrino mass is given below
\vspace{.5cm}
\begin{figure}[h]
\bcn
\epsfig{file=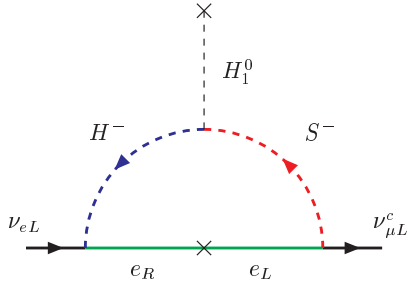,height=3cm}
\caption{Neutrino mass generation in the Zee model}
\label{fig:Zeemass}
\ecn
\end{figure}
The resulting mass matrix is symmetric and takes  the form
\beq
{\cal{M}}_{\nu}=m_o\left( \bray {ccc}0&a&b\\a&0&c\\b&c&0 \eray \right)
\label{eq:zeemass}
\eeq
where $m_0$ is an overall mass scale set by the mass of $S$. in Eq.(\ref{eq:zeemass})
$a,b {\mathrm and} c$ 
are functions of lepton masses and Higgs masses. The characteristics of the Zee mass
matrix are  the vanishing 
diagonal elements. This structure leads to
bi-maximal mixing \cite{bimax} and does not agree with the latest SNO data \cite{SNO}.
Furthermore the model suffers from being arbitrary. However, the mass matrix is intriguing
and can serve as the leading order to the correct mass matrix. To do so one has to find
a natural way of filling in some of the zero's in the Zee mass matrix.

This can be achieved in 5D unification models which possess gauge symmetries in the
bulk. The components of the the 5D fields that form representations of 
the symmetry can have different
 boundary values on the 4D brane. This reduces the symmetry that particles resides
on the brane sees. This
is known as orbifold symmetry breaking (OSB)mechanism.

 To appreciate this a little more we 
consider a 5D theory with the extra dimension compactified in a circle; i.e
 $z \in [-\pi R, \pi R]$. If we impose a parity of $Z_2$ such that a field $A(x,z)$
transform as 
\beq
P: z\longleftrightarrow -z  \;\;\; PA(z)=\pm A(z).
\eeq  
Clearly  $z=0$ and $z=\pi R$ are  fixed points and the background geometry is $S_1/Z_2$. Now we impose
a second parity by defining $z^{\prime}=z-\pi R/2$ such that
\beq 
P^{\prime}: z^{\prime}\leftrightarrow -z^{\prime}
\eeq
The fixed points are at $z=0$ and $z=\frac{\pi R}{2}$ and the geometry is 
$S_1/(Z_2\times Z_2^{\prime})$.
The fields can have different parities under $P$ and $P^{\prime}$. In particular the Fourier
decomposition of a bulk field into their  KK modes that conform to the parities
assignments are listed in Table I.
\begin{center}
\begin{tabular}{|c|c|c|c|c|}
  \hline
  $(P,P')$ &  form & mass & $z=0$ & $z=\frac{\pi R}{2}$ \\
  \hline
  $(++)$ & $\frac{1}{\sqrt{2 \pi R}}[{ A_0(x)} +\sqrt{2}\sum_{n=1} A^{++}_{2n}(x) \cos\frac{2nz}{R}] $
   & ${2n\over R}$ & $\surd$ & $\surd$ \\
  $(+-)$ & $\frpir[\sum_{n=0} A^{+-}_{2n+1}(x) \cos\frac{(2n+1)z}{R}] $
   & ${(2n+1)\over R}$ & $\surd$ & $X$ \\
  $(-+)$ & $\frpir[\sum_{n=0} A^{-+}_{2n+1}(x) \sin\frac{(2n+1)z}{R}] $
  & ${(2n+1)\over R}$  &$X$  & $\surd$ \\
  $(--)$ & $\frpir[\sum_{n=0} A^{--}_{2n+2}(x) \sin\frac{(2n+2)z}{R}] $
   & ${(2n+2)\over R}$ & $X$ & $X$ \\ \hline
\end{tabular}
\ecn
\bcn
Table I. KK decomposition of a bulk field $A(x,z)$ with parities $(P,P')$
\end{center}
The zero modes have $(+ +)$ parity and are identified as SM fields. As an example
consider $SU(5)$ with the Higgs field in the $\mathbf 5$ representation. We can assign
$(++)$ parities to the $SU(2)$ components and different parities to the $SU(3)$ 
color Higgs components.
The former is just the SM Higgs field and couples to SM matter fields which are placed 
on the $z=0$ brane
whereas the colored Higgs are KK excitations and will have masses $\frac{2n-1}{R}\ n=1 \dots $. 
They  latter will be
 heavy if the compactification radius $R$ is small and of the order of the unification scale.
This provide a  new natural solution to the
doublet-triplet splitting problem in $SU(5)$ GUT \cite{Kaw}. For a review of orbifold
GUT models see \cite{HN}.
With this technique we can now shape new tools for investigating the fermion mass problem.

In Ref.(\cite{CCN}) and (\cite{CN}) this has been applied to study neutrino mass 
generation without the benefit of $N_R$ fields for $SU(3)$ and $SU(5)$ unification 
models. The first model is particularly simple and illustrates the clearly physics 
involved and extends previous work on this model \cite{bsu3}. The $SU(3$ symmetry is in the
bulk and acts  on a background geometry of $S-1/Z_2\times Z_2^{\prime}$.
It only unifies the $SU(2)\times U(1)$ symmetry to $SU(3)_W$ and has
obvious built in lepton number number violation when the leptons are placed in the fundamental
${\mathbf 3}$ i.e. $(\nu_L \  e_L \ {e_R}^c)^T$. The unified value
of the weak mixing angle is well known to be given by
$\sin ^2 (\theta _W)={\frac14}$ \cite{SU3} and one can
expect unification to occur around $\sim {\mathrm {1-10\  TeV}}$ after renormalization 
group considerations. The lepton triplet is placed on
the $z=0$ brane and $SU(3)$ is broken by orbifolding the gauge fields and not by the Higgs
mechanism which is reserved for SM symmetry breaking. We require a bulk Higgs field in 
the $\mathbf{3}$ denoted by $\phi_3$  and a symmetric antisextet \sext 
to give realistic charged lepton masses. We denote fields in \sext by $\phi_6$. Due to
the requirement of $Z_2\times Z_2^{\prime}$ invariance we also introduce a second
Higgs triplet $\phi_3^{\prime}$ for the construction of the necessary \trip\sext\trip
coupling.

In this model the parities $P$ and $P^{\prime}$ are represented by $3\times 3^{\prime}$ matrices:
\beqa
P=\left(\begin{array}{ccc}
 1&0& 0  \\
 0 &1& 0 \\
 0& 0 & 1
\end{array}\right),  \hs
P^{\prime}=\left(\begin{array}{ccc}
 1&0& 0  \\
 0 &1& 0 \\
 0& 0 & -1
\end{array}\right).
\eeqa
The parities of the components of Higgs fields are engineered according to
phenomenological needs and are
\beqa
\phi_3=\left(\begin{array}{c}  \phi_3^-(++)\\\phi_3^0(++)\\ h_3^+(+-)
\end{array}\right),\hs
\phi'_3=\left(\begin{array}{c}  \phi_3^{'-}(+-)\\\phi_3^{'0}(+-)\\ h_3^{'+}(++)
\end{array}\right),
\label{P3}
\eeqa
and
\beqa
\label{P6}
\phi_6=\left(\begin{array}{ccc}
 \phi^{++}_{\{11\}}(+-) &\phi^{+}_{\{12\}}(+-)& \phi^{0}_{\{13\}}(++)  \\
 \phi^{+}_{\{12\}}(+-) &\phi^{0}_{\{22\}}(+-)& \phi^{-}_{\{23\}}(++) \\
 \phi^{0}_{\{31\}}(++) &\phi^{-}_{\{32\}}(++)& \phi^{--}_{\{33\}}(+-)
\end{array}\right).
\eeqa

With these ingredients 1 loop neutrino masses can be generated via
\vspace{1cm.}
\begin{figure}[h]
\bcn
\epsfig{file=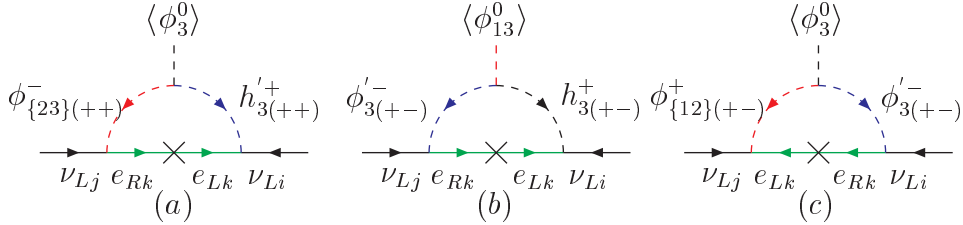,height=3cm}
\caption{Neutrino Mass Generation in SU(3) Model}
\label{fig:loopsu3}
\ecn
\end{figure}
The dominant contribution comes from \ref{fig:loopsu3}a and leads to a Zee-like neutrino
mass matrix. The next two diagrams give the necessary correction to this. In 
contrast to 4D models these terms exist as part of the 5D model. We do
not have to put them in by hand. Two parameters
central to all extra dimension models are the compactification radius $R$ and the
cutoff scale $M_*$ at which the field theory becomes strong. A lower limit on
$R^{-1}>1$TeV  can be obtained from the non observation of KK modes and $RM_* \sim 100$.
Without fine tuning of Yukawas other than that required to get the charged lepton
masses and using $R\sim$ 2 TeV one can obtain
\[
{\cal M}_\nu \sim
\left(\begin{array}{ccc}
  0.420 & 1.0 & 0.922 \\ 1.0 & 0.097 & -0.464 \\ 0.922& -0.464 &
  0.006 \end{array}\right)\times 0.0441 \,(\mbox{eV}).
\]
which  gives a good fit to the SuperK and SNO data. This solution is of the IMH
type and interestingly predicts a detectable neutrinoless double beta decay rate in the next 
round of experiments. Furthermore,
the  model has very rich phenomenology at the TeV scale which is the compactification
scale here. In particular double charged gauge boson are predicted for this model.
which can be searched for in linear colliders.
Rare decays such as $\mu \ra 3e$ and similar $\tau$ decays are also sensitive probes
of the model \cite{CN2}.

In the $SU(3)$ model it is known that quarks cannot be accommodated and has to be
placed at the other fixed point. It may be  more natural to place quarks and leptons on 
the same footing as in the $SU(5)$ model. To generate neutrino masses with 
only the fifteen  SM fermions
per family requires the use of either ${\mathbf {15}}$ or ${\mathbf {10}}$ bulk Higgs
field. The procedure is similar to the case of $SU(3)$ exploiting OSB. The addition
of exotic Higgs exacerbates the gauge unification problem in $SU(5)$. This can be
solved by additional fermions (see \cite{CN2}). 
After the dust settles unification is found to take place at $10^{15}$ GeV. An interesting
result emerges for this model. It is  more natural to get
the NMH  with the ${\mathbf 15}$ Higgs and IMH solutions can only be found with 
${\mathbf 10}$ Higgs fields.

\section{Conclusion}
 
We have reviewed the  construction of neutrino mass models in extra dimensions.
The possibility of large extra dimension; i.e $1/R \ll M_{\mathrm Planck}$ has resulted in
gaining very different perspectives in model building compared similar
exercises in 4D. These models also have
different low energy phenomenology than their 4D counterparts. In the Dirac neutrino case
one expects a sterile component in neutrino oscillation pattern. In the radiative mass
mechanism we expect KK modes of gauge bosons that couples to bileptons. The 4D models
on the other hand  have only exotic scalar particles. These models has a serious
down side. Thus far all of them  has many free
parameters and is no better or worse than the 4D models. We also encounter
very deep theoretical questions. What physics determines the
compactification scale remains unanswered. For the orbifold models the origin of 
parities of the various fields is a mystery. 

Due to the lack of space we have not discuss neutrino mass in models that implement
the seesaw mechanism in extra dimensions. This is studied for the $SO(10)$ case in
\cite{Raby}. Omitted are also discussions of neutrino masses in models
with warp extra dimensions
\cite{RS}. We will only mention that this has been explored in the case of Dirac neutrinos
\cite{GN} and \cite{McLN2}. Within the context of unification models the 
$SU(5)$ case was studied in \cite{wsu5}.
 
The many beautiful neutrino experiments have  given us the first hint of physics beyond
the SM. The bilarge neutrino mixing came as real surprise to theorists.
They have  stimulated us to explore scenarios which we have not ventured before. We can 
look forward for more data from the ongoing neutrino experiments and perhaps
even more unexpected results. 
Happily there will be more data to come both from high energy experiments with the anticipated
turning on of LHC as well as many precision measurements at lower energies.
Before the new phyiscs becomes clear it behooves us to keep an open
mind and be alert of new signatures.

I thank Prof C.W. Kim, Dr E.J. Park and the organizers of ICFP03
for their hospitality and for a stimulating conference.
I am grateful to Dr.W.F Chang for many discussions and for teaching me much during our
collaborations.

\end{document}